\documentclass[11pt,a4paper]{article}
\pdfoutput=1
\usepackage{jheppubmod}

\usepackage{braket}

\newcommand{\be}{\begin{equation}}
\newcommand{\ee}{\end{equation}}
\newcommand{\nn}{\nonumber}
\newcommand{\im}[1]{\text{Im}\left(#1\right)}

\begin{document}
\setcounter{tocdepth}{2}

%%%
%Title page
%%%

\title{Extremal Tunneling and Anti-de Sitter Instantons}
\author{Lars Aalsma,}
\author{Jan Pieter van der Schaar}
 
\affiliation{Institute for Theoretical Physics Amsterdam, Delta Institute for Theoretical Physics,\\
University of Amsterdam, Science Park 904, 1098 XH Amsterdam, The Netherlands}

\emailAdd{l.aalsma@uva.nl}
\emailAdd{j.p.vanderschaar@uva.nl}

 \abstract{We rederive and extend the amplitude for charged spherical shells tunneling through the outer horizon of charged black holes. In particular, we explicitly confirm that an effective action approach with natural initial conditions for a spherical shell, including backreaction, reduces to the tunneling integral. Consequently, we establish a universal expression for the probability of emission in terms of the change in the horizon entropy. Notably, the result for the charged black hole also captures the superradiant regime of charged particle decay at low energies. We then explore an appropriately regulated extremal and near-horizon limit, relating the tunneling amplitude to a family of gravitational instantons in the near-horizon Anti-de Sitter geometry, reducing to the known result for $AdS_2$ domain walls to leading order in the probe limit. We comment on the relation to the Weak Gravity Conjecture and the conjectured instability of (non-supersymmetric) Anti-de Sitter vacua.}
 
 \maketitle

%%%
% Main Body
%%%

\section{Introduction}

Ever since Hawking obtained his famous result for the thermal emission spectrum of black holes, an important question has been to understand, compute or estimate its leading corrections. The universal thermal nature of the spectrum is at the heart of the black hole information paradox and one unavoidable source of corrections is due to energy conservation: a black hole can only emit a particle with an energy at most equal to the mass of the black hole, implying the spectrum cannot be exactly thermal in any realistic micro-canonical description. 

In fact these backreaction corrections were first studied by Kraus and Wilczek by focusing on the dominant spherically symmetric sector of black hole emission \cite{Kraus1995a,Kraus1995}. They imposed energy conservation by constructing a (non-local) effective action for the spherical shell in which the (radial) gravitational degrees of freedom are integrated out. It was subsequently suggested by Parikh and Wilczek that these results could also be interpreted, and more easily computed, in terms of the amplitude of a single particle tunneling through the horizon \cite{Parikh2000}. The tunneling approach clearly points towards a universal answer for the probability which is always equal to the change in the black hole horizon entropy before and after emission, which was already pointed out in earlier work by Massar and Parentani \cite{Massar1997} using different methods. The fact that this probability is proportional to the change in the entropy supports the interpretation of the emission process, even after including backreaction, in terms of statistical thermodynamics \cite{Bekenstein:1975tw}.   

One motivation for this work was to better understand the relation between the effective action approach of Kraus and Wilczek and the tunneling approach of Parikh and Wilczek, over which there has been some confusion over the years. In particular, the final result of the original Kraus-Wilczek paper does not match the (universal) tunneling result, although this was apparently remedied in \cite{Keski-Vakkuri1997} for the case of neutral spherical shells emitted from a Schwarzschild black hole. We will show, in a more general charged black hole setting, that the approach of Kraus and Wilczek is indeed equivalent to the tunneling approach. As a corollary, this provides a thorough and `from first principles' effective action explanation for the validity of the tunneling approach, specified to the interesting case of charged particle shells. We will confirm that for a large range of parameters the probability for emission of charged spherical shells from a charged black hole indeed is proportional to 
\be \label{eq:UnivDecay}
P_{\omega,q} \propto e^{\Delta S_{BH}} ~,
\ee
as in the neutral spherical shell case. Here $\Delta S_{BH}=S(M-\omega,Q-q)-S(M,Q)$ is the change in entropy of the black hole before and after emission of a spherical shell with energy $\omega$ and charge $q$. Although we will be considering charged emission from a four-dimensional charged black hole, the appearance of the entropy difference and its associated interpretation in terms of statistical thermodynamics strongly suggests this result also applies to higher-dimensional black holes and/or black branes (in the spherically symmetric sector). 

After having carefully understood the detailed structure and universal nature of the result, we then study its implications in limits of interest. Specifically, we will show that the expression remains valid in the extremal limit of the black hole ($M=Q$) as long as the emitted particle shell satisfies $\omega \leq q$. The latter condition of course relates to the Weak Gravity Conjecture (WGC) \cite{Arkani-Hamed2007}, which essentially claims that in any consistent theory of quantum gravity there should exist a charged (elementary) particle whose mass is smaller than its charge in Planck units, i.e. $m \leq q$. This bound simply reflects the fact that an extremal black hole can only get rid of its charge by emitting a particle with $\omega \leq q$ to avoid creating a naked singularity. After all, one of the original motivations for the WGC was that an extremal black hole should be able to decay. We find that the probability to emit a charged particle satisfying the WGC from an extremal black hole is still nicely represented in terms of the entropy difference. To obtain a sensible result in the extremal limit crucially relies on including the backreaction of the shell. Moreover, we point out that the result remains applicable in a (non-thermal) regime of parameter space where the electrostatic potential energy dominates for some fixed particle charge $q$. At low enough energies the emission of particles of charge $q$ enters the so-called `superradiant' regime, where the entropy difference changes sign and the tunneling amplitude has to be reinterpreted. This low energy superradiant instability allows the black hole to quickly get rid of its charge, as originally noted and computed by Gibbons \cite{Gibbons1975}, but here we include the effect of backreaction. 

With the generalized result for charged emission from charged black holes at our disposal we will then study its consequences in relation to a conjectured extension of the WGC, as put forward in \cite{Ooguri2016,Freivogel2016}. The claim of these authors is that only supersymmetric BPS states, which saturate the WGC bound, can be (meta-)stable. If correct, this implies that not only (extremal) black holes, but also (non-supersymmetric) Anti-de Sitter vacua should feature universal decay channels. This conjectured extension of the WGC can be studied concretely in the context of an extremal Reissner-Nordstr\"om black hole, where the near-horizon geometry factorizes into $AdS_2$ times $S^2$. As we will show, the universal tunneling expression, when applied to the case of extremal to (non-)extremal emission, indeed implies a specific decay rate for non-extremal domain walls and at best meta-stability (fragmentation) for extremal domain walls in $AdS$ spacetimes. When we expand our near-horizon result to leading order in backreaction, it exactly reproduces the general $AdS_2$ Euclidean instanton action of Maldacena, Michelson and Strominger which describes an instability in the non-supersymmetric ($\omega < q$) case \cite{Maldacena1999}. In the extremal case, this result matches the instanton first discovered by Brill \cite{Brill1992}, corresponding to $AdS_2$ fragmentation. As shown in \cite{Maldacena1999}, in the limit where one of the charges is very small, the fragmentation amplitude indeed coincides with the Euclidean action of the Brill instanton, which we explicitly relate to the tunneling amplitude. As a consequence, our results, which fully incorporate backreaction in the spherically symmetric sector, confirm and extend the existence of a family of gravitational instantons describing the decay of $AdS$ vacua by the creation and subsequent expansion of super-extremal domain walls.

This paper is organized as follows. In section \ref{sec:UnivDecay} we employ the methods of Kraus and Wilczek to show that non-extremal charged black holes give rise to a universal decay rate of charged particles that is, after including backreaction, given by \eqref{eq:UnivDecay}. We demonstrate that this is equivalent to the tunneling prescription of Parikh and Wilczek. Continuing, we then carefully study the extremal limit of this result in section \ref{sec:extremal} and identify a `superradiant' region of parameter space where the charged emission is significantly enhanced as compared to the thermal regime. We go on to analyze the near-horizon limit of the tunneling calculation and interpret our results in terms of a family of gravitational instantons corresponding to instabilities of (non-supersymmetric) $AdS$ vacua. Finally, we discuss our results and present our conclusions in section \ref{sec:conclusion}. Some details on relevant integrals can be found in appendix \ref{app:integrals}.

\section{Inclusion of gravitational backreaction}\label{sec:UnivDecay}

In order to study corrections to Hawking radiation from backreaction, Kraus and Wilczek used an effective action to derive their results \cite{Kraus1995a,Kraus1995} whereas Parikh and Wilczek used a seemingly more ad-hoc approach by assuming particles tunneled through the horizon \cite{Parikh2000}. Obviously, since both approaches aim to incorporate the spherically symmetric part of the backreaction on the geometry of a black hole, the final result should be the same. However, on a technical level these approaches seem to be rather different and the expected agreement is far from obvious. Notably, the supplied boundary conditions, which are related to the physical interpretation, are different in the two cases. Despite these differences, we will in this section verify that the results are the same, and we will clarify some of the sources of confusion.  

In order to include the effects of the energy of the spherical shell in the emission process one can either fix the total energy of the spacetime or the black hole mass. While Kraus and Wilczek fix the black hole geometry and let the ADM mass vary, Parikh and Wilczek fix the ADM mass and allow the black hole geometry to fluctuate. Because black hole evaporation corresponds to the loss of black hole mass-energy to the asymptotically flat space surroundings, one might be inclined to prefer the Parikh-Wilczek approach. However, we would like to emphasize that the effective action approach of Kraus and Wilczek, which might be considered a more rigorous derivation of the spherically symmetric dynamics including backreaction, can just as well be applied with the ADM mass fixed. As a consequence, one should be able to derive the (universal) Parikh-Wilczek tunneling answer from a first principles effective action method.

To illustrate this, we will first derive the decay rate of a charged spherical shell from a charged black hole using the Kraus-Wilczek effective action method and subsequently employ the tunneling perspective to arrive at the same result more directly. Along the way we will show how the Kraus-Wilczek computation \cite{Kraus1995a,Kraus1995} reduces to the one performed by Parikh and Wilczek \cite{Parikh2000}.

\subsection{The effective action of a spherical shell}

The central idea of the Kraus-Wilczek approach is that the dominant contribution to the emission flux in Hawking's original calculation is in the s-wave sector (spherical shells). Even though backreaction is in general hard, if not impossible, to keep track of, focusing on just the s-wave contribution allows backreaction to be incorporated. By constraining the gravitational degrees of freedom one arrives at a two-dimensional (non-local) effective action for a spherical shell in the black hole background. Of particular importance are the boundary conditions that are needed to explicitly determine the on-shell action, which is then used in a WKB approximation to construct solutions to the (corrected) field equations. Using these corrected mode functions, the overlap can then be computed between appropriate energy eigenstates in terms of asymptotic Minkowski time and the Unruh vacuum state, which is selected by a specific initial condition for the mode functions near the horizon. Following through, one then arrives at Hawking's result plus corrections due to gravitational backreaction in the s-wave sector. In the original article \cite{Kraus1995a} this was done for neutral emission from a Schwarzschild black hole, and in a follow-up article \cite{Kraus1995} the authors report to have worked out the result for charged emission as well. We will summarize the computation in the more general case of charged emission below and show that the corrected final result agrees with the elegant and universal answer that is naturally obtained and understood from a perspective of particle tunneling. 

Lets us start by considering a four-dimensional Reissner-Nordstr\"om black hole with mass $M$ and electric charge $Q$ with the metric and gauge field $A$ given by
\begin{align}
ds^2 &= -f(r)dt^2 + f(r)^{-1} dr^2 + r^2 d\Omega_2^2 ~, \label{eq:stdmetric} \nn\\
f(r) &= 1-\frac{2M}{r} + \frac{Q^2}{r^2} ~, \nn\\
A &= -\frac{Q}{r} dt ~.
\end{align}
We introduce the standard notation for the inner and outer horizon of the black hole.
\be
r_\pm = M \pm \sqrt{M^2-Q^2} ~.
\ee
The metric \eqref{eq:stdmetric} contains a coordinate singularity at $r_{\pm}$, so in order to construct regular mode functions for a freely falling observer we introduce coordinates that are regular across the horizon. A particularly useful choice are the Painlev\'e-Gullstrand coordinates \cite{painleve1921cr}. We define a new time coordinate as
\be
t_p = t + g(r) \,, \qquad g(r) = \int dr\, \frac{\sqrt{1-f(r)}}{f(r)} ~,
\ee
such that the metric becomes (dropping the subscript $p$)
\be \label{eq:painleve}
ds^2 = -f(r)d t^2 + 2\sqrt{1-f(r)}dt dr + dr^2 + r^2 d\Omega_2 ~,
\ee
which is regular at the horizon.

\subsubsection{Quantization}

Now that we have specified the details of the background, we turn to quantization of (spherically symmetric) fields in this background. Since we are interested in charged radiation, we consider a complex scalar field $\phi(t,r)$ and write down its mode expansion. Considering modes that are positive frequency with respect to the Killing time that is used by an asymptotic observer we write
\be
\phi(t,r) = \int dk \, \left( \hat c_k u_k^q(r) e^{-i\omega_k t} + \hat d_k^\dagger \bar {u}_k^{-q}(r) e^{+i\omega_k t} \right) ~.
\ee
Here $u_k^q(r)$ denotes a particle mode function with positive charge $q$ and $\bar{u}^{-q}_k(r)$ an anti-particle mode function with negative charge $-q$. The bar indicates complex conjugation. Furthermore, $k$ is the wavenumber with $\omega_k$ the corresponding energy. We can now define the vacuum of an asymptotic observer as
\be
\hat c_k \ket{0_A} = \hat d_k \ket{0_A}=0 ~.
\ee
Alternatively, we can expand the scalar field in a different set of modes that are positive frequency with respect to a freely falling observer as
\be
\phi(t,r) = \int dk \, \left( \hat a_k v_k^q (t,r) + \hat b_k^\dagger \bar{v}_k^{-q}(t,r) \right) ~,
\ee
and define the vacuum of a freely falling (Unruh) observer as
\be
\hat a_k \ket{0_U} = \hat b_k \ket{0_U}=0 ~.
\ee
The two sets of creation and annihilation operators are related by the following Bogoliubov transformations
\begin{align}
\hat c_{k} &= \int dk'\left(\alpha_{kk'}\hat a_k + \beta_{kk'}\hat b_{k}^\dagger\right) ~,\\
\hat d_{k}^\dagger &= \int dk'\left(\bar \alpha_{kk'}\hat b_k^\dagger + \bar\beta_{kk'}\hat a_k\right) ~.
\end{align}
The Bogoliubov coefficients can be expressed in terms of the following integrals
\begin{align}
\alpha_{kk'} &= \frac{1}{2\pi u_k^q(r)}\int_{-\infty}^\infty dt \, e^{i\omega_{k'} t}v^q_{k}(t,r) ~, \\
\beta_{kk'} &= \frac{1}{2\pi u_k^q(r)}\int_{-\infty}^\infty dt \, e^{i\omega_{k'} t}\bar{v}^{-q}_{k}(t,r) ~,
\end{align}
which satisfy the standard orthonormality and completeness constraints.
Selecting the Unruh vacuum state, the amplitude for detecting $n$ particles (and $n$ anti-particles) with momentum $k$ is determined by the overlap  
\be
\Gamma^n_{k} = \left< 0_U | n_k^q, n_{-k}^{-q} \right> \, . 
\ee
The average number of particles with momentum $k$ in the Unruh state, introducing the number operator $\hat{N}_k \equiv \hat{c}_k^\dagger \hat{c}_k$, is given by 
\be 
\bra{0_U} \hat{N}_k \ket{0_U} = \int dk^\prime |\beta_{kk'}|^2 \, . 
\ee 
Assuming the different mode expansions are defined on the same spatial slices, as will be our case of interest, the Bogoliubov matrices will be diagonal and the $k'$ index can be dropped. Integrating over all modes and appropriately regulating the expression (by introducing a finite space-time volume) one arrives at the result for the (average) total flux of asymptotically observed particles. The integrand, corresponding to the average flux density with an energy between $\omega_k$ and $\omega_k + d\omega_k$, equals 
\be
F(\omega_k) = \frac{d\omega_k}{2\pi} \frac{\Omega(\omega_k)}{|\alpha_{k}|^2/|\beta_{k}|^2 - 1} \,.
\ee
Here an additional grey-body factor $\Omega(\omega_k)$ was introduced that describes the effects of re-scattering off the potential. In the case that we would ignore backreaction, assuming the emission were exactly thermal and the background perfectly transparent ($\Omega(\omega_k) = 1$), the ratio $|\beta_{k}|^2/|\alpha_{k}|^2$ would equal the Boltzmann factor characterizing the Bose-Einstein distribution. Throughout this paper we will ignore the effects of a non-trivial grey-body factor, implying that the probability for the black hole to emit a single quantum with momentum $k$ equals   
\be
P_k = | \Gamma^1_k |^2 = \frac{|\beta_k|^2}{(1+|\beta_k|^2)^2} = \frac{1}{|\alpha_k|^2} \frac{|\beta_{k}|^2}{|\alpha_{k}|^2} \,, 
\ee
where in the final equality we used the normalization $|\alpha_k|^2 - |\beta_k|^2=1$ which, as we will explain later, cannot be assumed for charged particle emission at low enough energies. For a thermal distribution this probability is of course proportional to the Boltzmann factor. 

In principal, we need to know the explicit form of the mode functions $v_k(t,r)$ in order to calculate the Bogoliubov coefficients, which is not straightforward. However, as was argued by \cite{Kraus1995a,Kraus1995}, the mode functions take on a simple form in the WKB approximation, which is valid for short wavelengths. Because modes near the horizon are infinitely blueshifted with respect to an asymptotic observer, this approximation should be valid as long as the modes are close enough to the horizon, which is then sufficient to determine the Bogoliubov coefficients describing the emission process. Thus, the mode functions are assumed to be of the following WKB form
\be
v_k^q(t,r) = e^{i S_k^q(t,r)} \,, \qquad \bar{v}^{-q}_k(t,r) = e^{-i S^{-q}_k(t,r)} ~.
\ee
Here, $S^{\pm q}_k(t,r)$ is the classical action of the shell and the superscript $\pm q$ indicates the charge of the solution. To obtain an explicit and useful expression for the effective action, we will make use of a Hamiltonian formalism.

\subsubsection{Effective action from a Hamiltonian formalism}

In \cite{Kraus1995a,Kraus1995} a Hamiltonian formalism is used to derive the effective action of a particle in the s-wave approximation, i.e. describing the dynamics of a shell in a black hole background incorporating the backreaction of the shell. As is well known, in order for the variation of the gravitational action to vanish when evaluated on the equations of motion, it is necessary to supplement the action with boundary terms that cancel the ones that are induced by the variation, see e.g. \cite{Hawking1996}. For non-extremal black holes in asymptotically flat spacetimes there are two types of surface terms that require cancellation. One of these is defined asymptotically and yields the ADM mass of the spacetime, whereas the second one is defined on the black hole horizon and is related to its area and surface gravity \cite{Massar1998}. 

In \cite{Kraus1995a,Kraus1995} the geometry of the black hole is kept fixed and the ADM mass is allowed to vary to satisfy the Hamiltonian constraints. 
This means that the boundary term on the horizon vanishes and we only need to subtract the asymptotic boundary term from the action to have a well-defined variational principle. 
However, we could also have fixed the ADM mass as was done in \cite{Parikh2000}. In this case, we should add the boundary term defined on the horizon to the action. 

For all practical purposes, this means that in the Kraus-Wilczek approach the evolution of the shell is determined by the ADM mass and total charge of the system (and therefore the geometry {\it outside} the shell), whereas in the Parikh-Wilczek method it is the mass and charge of the black hole that determines the evolution (the geometry {\it inside} the shell). This is precisely the difference in boundary conditions that we alluded to before. For the purpose of consistently comparing with \cite{Kraus1995} we will fix the black hole geometry for now, but it should be stressed that this is just a choice and we could just as well have fixed the ADM mass.

By solving the Hamiltonian constraints, it was found in \cite{Kraus1995} that the introduction of a massless shell with energy $\omega$ and charge $q$ splits the spacetime into two parts with mass parameter ${\cal M}(r)$, which appears in the metric as $f(r)=1 -2{\cal M}(r)/r$.
\begin{align}
{\cal M}(r) =
\begin{cases}
M - \frac{Q^2}{2r} & (r<\hat r) ~,\\
\\
M+\omega - \frac{(Q+q)^2}{2r} & (r>\hat r)  ~.
 \end{cases}
\end{align}
Here, $\hat r$ is the position of the shell. The classical action of the shell can be written as \cite{Kraus1995a,Kraus1995}
\be \label{eq:canaction}
S_k^q(t, r(t)) = S_k^q(0, r(0)) + \int_{r(0)}^{r(t)} d r \, p_c - (M_+ - M)t  ~.
\ee
In this expression, $r(t)$ is trajectory of the shell, $p_c$ its canonical momentum whose explicit form is given in \eqref{eq:canmom} and $M_+$ the ADM mass of the spacetime. We subtracted the contribution of the black hole from the action such that the Hamiltonian is that of the shell: $(M_+-M)=\omega$. From Hamilton's equations we find that the equation of motion of the shell is
\be \label{eq:outgoingEOM}
\dot r = \frac{\partial H}{\partial p_c} = 1 - \sqrt{\frac{2(M+\omega)}{r}-\frac{(Q+q)^2}{r^2}} ~.
\ee
This is the equation of motion of an outgoing null geodesic in a spacetime with a black hole of mass $M+\omega$ and charge $Q+q$, as can be seen by solving
\be
g_{\mu\nu}\frac{dx^\mu}{dt}\frac{dx^\nu}{dt} = 0 ~,
\ee
in Painlev\'e-Gullstrand coordinates. This indeed agrees with our earlier observation that the imposed boundary conditions on the ADM mass and total charge should determine the evolution of the shell. 

To find an explicit expression for the trajectory $r(t)$, we need to specify the initial position $r(t=0)$. The natural choice is to demand the standard positive and negative frequency modes for a freely falling observer that crosses the horizon (the Unruh vacuum), i.e. we impose that the mode functions take the form of a standard plane wave at $t=0$:
\be \label{eq:initcond}
S_k^q(0,r(0)) = kr(0) ~.
\ee
This means that the (diagonal) Bogoliubov coefficients can be written as
\begin{align}
\alpha_{k} &= \frac{1}{2\pi u_k^q(r)}\int_{-\infty}^\infty dt \, e^{i\omega_{k} t + iS_k^q(t,r)} ~,\\
\beta_{k} &= \frac{1}{2\pi u_k^q(r)}\int_{-\infty}^\infty dt \, e^{i\omega_{k} t - iS_k^{-q}(t,r)} ~.
\end{align}
We can compute these integrals by 
 a saddle point approximation. The saddle points of these integrals are solutions to
\be \label{eq:saddleeq}
\omega_{k} \pm \frac{\partial S_{k}^{\pm q}(t,r)}{\partial t} = 0 ~,
\ee
where the plus sign corresponds to $\alpha_{k}$ and the minus sign to $\beta_{k}$. This simply indicates that the different saddle point trajectories have opposite energy. Because $\partial S^q_k/\partial t$ is minus the Hamiltonian we find that the solution to \eqref{eq:saddleeq} is given by
\be \label{eq:saddle}
\quad M_+ = M \pm \omega_{k} ~,
\ee
and the Bogoliubov coefficients at the saddle points are given by
\begin{align}
\alpha_{k} &\propto \exp \left( i k r(0) + i \int_{r(0)}^{r(t)} dr\, p_c \right) ~, \\
\beta_{k} &\propto \exp \left( -i k r(0) - i \int_{r(0)}^{r(t)} dr\, p_c \right) ~.
\end{align}
We will evaluate these expressions as close as possible to the horizon to minimize corrections to the WKB approximation. At the same time we should remain slightly outside the horizon for the mode functions $u_k^q(r)$ to be regular, so we take $r(t)$ to be just outside of the horizon, i.e. $r(t)=r_+(M+\omega,Q+q) + \epsilon$ with $\epsilon \ll r_+(M+\omega,Q+q)$. Before we can evaluate the Bogoliubov coefficients, we will need some explicit details of the trajectory of the shell towards which we will move our attention next.

\subsubsection{Evaluating the Bogoliubov coefficients}

The canonical momentum of an outgoing shell in our background is given by \cite{Kraus1995}
\be \label{eq:canmom}
p_c(r(t)) = \sqrt{2M r-Q^2} - \sqrt{2M_\pm r-Q_\pm^2} -  r \log\left(\frac{r-\sqrt{2M_\pm r-Q_\pm^2}}{r-\sqrt{2Mr-Q^2}}\right) ~.
\ee
The upper sign of $M_\pm,Q_\pm$ denotes the saddle point for $\alpha_k$ and the lower sign the saddle point for $\beta_k$. Null geodesics can be found by introducing lightcone coordinates $v$ and $u$ as
\begin{align}
v &= t + r_* = \text{constant}~,  \nn \\
u &= t - r_* = \text{constant} ~,
\end{align}
where we introduced the tortoise coordinate $r_*$.
\be
r_* = \int dr\, \frac{1}{f(r)} = r + \frac{r_+^2}{r_+-r_-}\log\left(r-r_+\right) - \frac{r_-^2}{r_+-r_-}\log\left(r-r_-\right)  ~.
\ee
It is now straightforward to obtain an explicit expression for $t$ and $k$, given the initial position $r(0)$ of the shell. By setting $u(t,r(t)) = u(0,r(0))$ it follows that
\be \label{eq:timeexpr}
t= r - r(0) +\frac{r_+^2}{r_+-r_-}\log\left(\frac{r-r_+}{r(0)-r_+}\right) - \frac{r_-^2}{r_+-r_-}\log\left(\frac{r-r_-}{r(0)-r_-}\right) ~.
\ee
Using the initial condition \eqref{eq:initcond} we obtain an expression for $k$.
\be \label{eq:kexpr}
k =p_c(0,r(0)) ~.
\ee
The last details we need are the value of $r(0)$ and $t$ at the saddle points. Because the modes are infinitely blueshifted near the horizon as compared to an asymptotic observer, we only require the solution in the limit $k\to \infty$. In this limit we can invert \eqref{eq:kexpr} to find an expression for $r(0)$ and plug this into \eqref{eq:timeexpr} to obtain an expression for $t$. This then leads to
\begin{align} \label{eq:saddletime}
r(0) &= 
\begin{cases}
r_+(M_+,Q_+) + {\cal O}(e^{-k/r_+})   & (\text{for } \alpha_{kk'}) \\
 \\
r_+(M_-,Q_-) - {\cal O}(e^{-k/r_+})   &(\text{for } \beta_{kk'}) \\
\end{cases}
 \\
\nonumber \\
\im{t}&=
\begin{cases}
0 &  (\text{for } \alpha_{kk'}) \\
 \\
-\pi\frac{r_+(M_-,Q_-)^2}{r_+(M_-,Q_-)-r_-(M_-,Q_-)}  &  (\text{for } \beta_{kk'}) ~. \\
\end{cases}
\end{align}
It is important to notice that generically, in the parameter space of the shell spanned by $q$ and $\omega$, $r(0)$ lies outside of the black hole horizon for $\alpha_{k}$. As a consequence, the action at the saddle point of $\alpha_{k}$ is completely real, and describes a classically allowed trajectory. On the other hand, the initial position of the shell is inside the horizon for $\beta_{k}$. This implies that the shell travels on a classically forbidden trajectory and the action picks up an imaginary part. Since the quantity of interest is the ratio of the absolute values of the Bogoliubov coefficients, implying that only the imaginary parts of the action contribute, we arrive at the following result
\be \label{eq:bogoratio}
\frac{|\beta_{k}|^2}{|\alpha_{k}|^2} \propto \exp\left[2\,\im{\int_{r(0)}^{r(t)} dr \, p_c} \right] ~,
\ee
which is evaluated at the saddlepoint for $\beta_{k}$ and where $r(t)$ is taken as close as possible, but slightly outside the horizon. Details on how to evaluate this integral and the correct pole prescription can be found in appendix \ref{app:integrals}. Here, we simply quote the result:
\be \label{eq:KWint}
\im{\int_{r(0)}^{r(t)} dr ~ p_c } = -\pi \int_{r_+(M_-,Q_-)}^{r_+(M,Q)} dr\, r   = \frac12 \pi (r_+(M_-,Q_-)^2 - r_+(M,Q)^2) \,.
\ee
This expression equals half the difference of the black hole entropy before and after emission, which is typically negative and reduces to the Boltzmann factor in the limit where the backreaction can be neglected. We therefore conclude that the probability of the black hole to emit a particle with charge $q$ and energy $\omega$, assuming $\Delta S_{BH}$ is negative, equals 
\be \label{eq:finalresult}
P(k) \propto \frac{|\beta_{k}|^2}{|\alpha_{k}|^2} \propto e^{\pi(r_+(M_-,Q_-)^2 - r_+(M,Q)^2)}= e^{\Delta S_{BH}} \,,
\ee
where $\Delta S_{BH} = S_{BH}(M-\omega,Q-q) - S_{BH}(M,Q) $. 

For both the neutral and charged case this does not exactly reproduce the result of the original articles \cite{Kraus1995a,Kraus1995} due to a technical error, which was in fact corrected in \cite{Keski-Vakkuri1997} for the neutral case. Here we extended it to also include charged emission. The fact that the shell (generically) follows a classically forbidden trajectory clearly suggests that we should be able to reproduce the result \eqref{eq:finalresult} directly by doing a tunneling calculation, an idea that was worked out in \cite{Parikh2000}. In the next subsection we demonstrate that this approach is indeed equivalent and verify explicitly that the same Hamiltonian formalism used in \cite{Kraus1995a,Kraus1995} underlies this computation.

\subsection{The tunneling perspective}\label{sec:tunnelmethod}

In the previous section we saw that the computation of the Bogoliubov coefficients reduced to calculating the imaginary part of the classical action, which is what one would compute in a tunneling calculation in a WKB approximation. This was done in \cite{Parikh2000} for the emission of neutral massless radiation. Here we generalize their calculation to charged emission and make clear that this method is equivalent to, and can be derived from, the Kraus-Wilczek effective action approach. 

As mentioned, an important difference between the two approaches is that the former keeps the black hole mass fixed and allows the ADM mass to vary, while the latter keeps the ADM mass fixed and allows the black hole mass to vary. Before we continue to discuss the tunneling calculation of Parikh and Wilczek, we first discuss some details of the effective action computation if one would have fixed the ADM mass. In that case the mass parameter ${\cal M}(r)$ for a shell at position $\hat r$ is given by
\begin{align}
{\cal M}(r) =
\begin{cases}
M-\omega - \frac{(Q-q)^2}{2r} & (r<\hat r)  ~, \\
 \\
M - \frac{Q^2}{2r} & (r>\hat r)  ~.
 \end{cases}
\end{align}
Since we fixed the ADM mass $M$, it is now the geometry inside the shell that determines the evolution. Again, we could use Hamilton's equations to obtain the result that the shell travels on a null geodesic with mass parameter ${\cal M}(r<\hat r)$ as the Hamiltonian of the shell is still given by $M_{ADM}-(M-\omega)=\omega$. Another difference is related to the initial condition of the shell which is now given by
\begin{align}\label{eq:Mfixedboundary}
r(0)=
\begin{cases}
r_+(M,Q) + {\cal O}(e^{-k/r_+}) &  (\text{for } \alpha_{k}) ~, \\
& \\
r_+(M,Q) - {\cal O}(e^{-k/r_+}) &  (\text{for } \beta_{k}) ~.  \\
\end{cases}
\end{align}
So for $\alpha_{k}$ the shell now starts just outside of, and for $\beta_{k}$ just inside of the \emph{initial} horizon. It is important to notice that the parameter $M$ that now appears in all relevant expressions is the ADM mass and not the black hole mass. At the end of the day, we want to interpret the probability for shell emission in terms of the change in entropy of the black hole. Hence, we should write the canonical momentum in terms of $M_{BH} = M_{ADM} - \omega$. After this simple shift, the calculation becomes equivalent to the Kraus-Wilczek computation with the black hole mass fixed. We therefore conclude that when fixing the ADM mass the Kraus-Wilczek effective action approach also leads to the same result \eqref{eq:finalresult}.

In addition, this result can now be compared directly with the tunneling method of \cite{Parikh2000}. The starting point of Parikh and Wilczek is the fact that in a WKB approximation the tunneling probability is given by the exponential of the classical action, which reduces to the integral 
\be \label{eq:tunnelingaction}
P \propto \exp\left[-2~ \im{ \int_{r_i}^{r_f} dr ~ p_c} \right] ~,
\ee
where $r_i$ and $r_f$ correspond to the initial and final position respectively of the particle that is tunneling through a potential barrier. Based on the previous section, we recognize it as the final expression for the integral in the Kraus-Wilczek approach. In fact, it can be directly related to the expression \eqref{eq:bogoratio} obtained by fixing $M_{ADM}$ and using the boundary conditions \eqref{eq:Mfixedboundary}. The relative minus sign between these expressions can be explained by the fact that in the tunneling integral \eqref{eq:tunnelingaction} $r_f < r_i$, since the shell is taken to tunnel from just inside the initial horizon to just outside the final horizon. In contrast, in the Kraus-Wilczek computation $r(0)$ is always smaller than $r(t)$, when expressed in terms of the black hole mass. Because these expressions are written in terms of the canonical momentum and do not (explicitly) depend on the details of the background, we expect this result to remain universally valid as long as spherical symmetry is imposed.

Making use of Hamilton's equations, we can manipulate \eqref{eq:tunnelingaction} to write it as
\be \label{eq:TunnelingAction}
\im{ \int_{r_i}^{r_f} dr \, p_c} = \im{\int_{r_i}^{r_f} dr \int_{H(0)}^{H(\omega)} dH \, \frac{1}{\dot r} } ~.
\ee
Here $H$ is the Hamiltonian of the geometry seen by the shell. Since the shell follows a null geodesic in a geometry with mass $M-\omega$ and charge $Q-q$, the Hamiltonian is identified as the mass of the black hole, such that $dH = -d\omega$. The equation of motion for the outgoing positive energy shell in Painlev\'e-Gullstrand coordinates is given by
\be
\dot r = 1 -\sqrt{\frac{2(M-\omega)}{r} -\frac{(Q-q)^2}{r^2}} ~.
\ee
The boundaries of the integral are taken such that we integrate the shell from just inside the initial horizon to just outside the final horizon. We now find that the integral of \eqref{eq:TunnelingAction} contains a pole, determined by position of the outer horizon, i.e. $r_+(M-\omega,Q-q)$. In order to evaluate this integral, we need a prescription that tells us how to deform the contour around the pole. Different choices correspond to different boundary conditions. We show in appendix \ref{app:integrals} that the prescription that supplies the (physically) correct boundary conditions is given by the (Feynman) deformation $\omega \to \omega - i\epsilon$, which was also used in \cite{Parikh2000}.

Evaluating the integral using the prescribed contour deformation and taking the boundaries as the position of the initial and final horizon, one arrives at
\be \label{eq:imaction}
\im{\int_{r_i}^{r_f} d r~ p_c} = -\pi \int_{r_+(M,Q)}^{r_+(M_-,Q_-)} dr ~ r = -\frac12\pi\left(r_+(M_-,Q_-)^2 - r_+(M,Q)^2\right) ~,
\ee
where we used the results of appendix \ref{app:integrals}. The fact that these are the correct boundaries to take can be seen by switching the order of integration, which leads to the same result \cite{Parikh2000}. So we see that the tunneling method indeed gives the same result, as it should, for generic parameters $\omega$ and $q$ implying the following universal decay probability (for sufficiently large energies $\omega$) 
\be \label{eq:decayrate}
P(k) \propto e^{\Delta S_{BH}} ~, 
\ee
where we (again) ignored the appropriate normalization factor, which for large enough negative values of $\Delta S_{BH}$ is approximately one. This universal expression can now be employed to study different physical scenarios. In \cite{Parikh2000}, where neutral radiation was considered, the result was used to identify the (leading) correction to Hawking radiation, capturing a deviation from perfectly thermal behavior, but consistent with an interpretation in terms of statistical thermodynamics. We will instead use this generalized expression to study charged decay channels in certain limits of parameter space that are of interest to us and where the inclusion of backreaction is crucial. We will in particular be considering limits where the emission of charged quanta does not (only) occur through a thermal Hawking process, but is dominated by a charged Schwinger-like process.

\subsection{Superradiant emission and the tunneling integral}

A particular limit of interest is that of low-energy charged emission, which is well known to display superradiant behavior. Indeed, in the regime of parameters where one expects superradiance the entropy difference $\Delta S_{BH}$ becomes positive, implying that the standard interpretation in terms of a tunneling probability is invalid. The appearance of a superradiant regime in the (incorrect) expression for the emission probability including backreaction was noticed in \cite{Kraus1995}, but not elaborated upon. Here we will provide the appropriate interpretation and application of the tunneling integral in the low-energy superradiant regime.

To remind the reader, usually superradiance is associated to (and described by) a scattering process, and as a consequence one introduces transmission and reflection coefficients instead of Bogoliubov coefficients. When scattering an incoming wave $v_{1,k}(t,r)$ on the horizon, conservation of flux implies
\be
v_{1,k} + R~v_{2,k} = T~ v_{3,k} ~,
\ee
where $v_1$ and $v_2$ are respectively the right-moving and left-moving wave functions inside the horizon and $v_3$ is the right-moving wave outside the horizon. The reflection and transmission coefficients $R$ and $T$ are normalized as
\be \label{eq:normalis}
|R|^2 + |T|^2 =  1 ~.
\ee
It is then straightforward to show that the transmission and reflection coefficients can be related to the Bogoliubov coefficients in the following way \cite{Hansen1981}
\be
|\alpha_{k}|^2 = 1/|R|^2 ~, \qquad \frac{|\beta_{k}|^2}{|\alpha_{k}|^2} = |T|^2 ~,
\ee 
where the standard normalization condition for the Bogoliubov coefficients has been assumed
\be
|\alpha_{k}|^2 - |\beta_{k}|^2 = 1 ~.
\ee
We conclude that the transmission coefficient $T$ can be associated to the tunneling probability $P$, which as we have seen is expressed in terms of the entropy difference between the final and initial state of the black hole. 

Obviously an interpretation in terms of a probability requires the ratio of Bogoliubov coefficients to be smaller than one. For charged emission there exists a parameter regime at low enough energies where the sign of $\Delta S_{BH}$ actually becomes positive. For charged decay channels obeying
\be \label{eq:superrad}
\omega + \frac{q^2}{2r_+} < q \frac{Q}{r_+} ~,
\ee
the change in entropy becomes positive and the tunneling integral is exponentially enhanced instead of suppressed. We recognize the left hand side as the total energy of the shell (including its electromagnetic self-energy) and the right hand side as the electromagnetic potential of the black hole that the shell couples to. Notably, in the extremal limit $M=Q$, the effect of backreaction is to shift the superradiant regime to lower (super-extremal $\omega<q$) values for the energy of the emitted particle. When \eqref{eq:superrad} is satisfied this clearly signals a (thermodynamic) instability, as it becomes possible for the black hole to radiate away charge, while nevertheless increasing the entropy of the black hole. This superradiant instability was first discovered in the process of partial wave scattering off rotating black holes. For rotating black holes it is absent in the s-wave sector (and therefore more suppressed), but for charged black holes it remains present when restricting to the s-wave sector at low enough energies. In the appropriate superradiant scattering process, the normalization condition for reflection and transmission coefficients for a particle with frequency $\nu$ and charge $q$ is affected in the following way \cite{Brito2015}.
\be
|R|^2  = 1 - \frac{\nu - q Q/r_+}{\nu}|{\cal T}|^2
\ee
So effectively this corresponds to the replacement   
\be
|T|^2 \to \frac{\nu - q Q/r_+}{\nu}|{\cal T}|^2 ~.
\ee
Comparing this to \eqref{eq:normalis} we observe that when the frequency obeys the bound
\be
\nu < q \frac{Q}{r_+} ~,
\ee
the reflection coefficient exceeds unity, i.e. $|R|^2>1$, meaning that the particle that scatters off the black hole takes away some of its mass and charge \cite{Brito2015}. This frequency agrees with \eqref{eq:superrad} in the limit $q/2Q \ll 1$, i.e. when ignoring backreaction. However, instead of particles scattering off black holes, we would like to consider spontaneous emission in this superradiant regime of parameter space. One observes that in the superradiant regime apparently $|\alpha_k|^2 = 1/|R|^2 < 1$, suggesting that the Bogoliubov coefficients should be interchanged ($\alpha_k \leftrightarrow \beta_k$) to still obey the normalization condition. Equivalently, one can interpret this as a change in the sign of the normalization condition for the Bogoliubov coefficients. This can be traced back to the fact that what was previously defined to be a positive frequency mode at asymptotic infinity in the superradiant regime turns into a negative frequency mode, and vice-versa. As a consequence, in the superradiant regime the probability of emission $P(k)$ for a charged particle should be re-evaluated and is related in a more indirect way to the tunneling integral, as we will see below. 
  
To determine the probability distribution we start with the appropriate expression for the average number of particles per mode in the superradiant regime \cite{Bekenstein:1977mv}. This is most easily derived by applying a change of sign for the normalization of the Bogoliubov coefficients, i.e. what one means with positive and negative frequency modes. This results in the following expression for the expectation value of the number operator in the superradiant regime 
\be 
\left< {N_k} \right> = \frac{-1}{|\alpha_{k}|^2/|\beta_{k}|^2 - 1} ~.
\ee
The change of sign in the numerator ensures that the average number of particles remains positive in the superradiant regime where the ratio $|\alpha_{k}|^2/|\beta_{k}|^2 = e^{-\Delta S_{BH}(k)} < 1$. It is important to note that at the transition from superradiant to ordinary (Hawking) emission it is crucial to take into account the greybody factor $\Omega(\omega_k)$ to ensure appropriately continuous behavior, but for our purposes here we can safely ignore this issue. The relevant probability distribution for observing $n$ particles in a mode $k$ can be written in terms of the average number as follows 
\be
P_k(n) = \frac{\braket{N_k}^n}{(\braket{N_k}+1)^{n+1}} ~.
\ee
As an easy check this indeed reproduces the standard (Bose-Einstein) distribution for single particle emission when $\Delta S_{BH}$ is negative. Using the superradiant expression for the average number of particles, one then arrives at the following probability for emitting a single particle in mode $k$
\be
P(k)_{SR} = (1-e^{-\Delta S_{BH}(k)})\frac{1}{(2-e^{-\Delta S_{BH}(k)})^2} ~, 
\ee
where we expressed the probability explicitly in terms of the ratio $|\alpha_{k}|^2/|\beta_{k}|^2 = e^{-\Delta S_{BH}(k)} < 1$. This superradiant expression clearly differs from the standard (Bose-Einstein) distribution and generalizes the known result without backreaction \cite{Bekenstein:1977mv}. Typically $\Delta S_{BH}>0$ over a considerable range of superradiant frequencies and the probability distribution is very flat. As a consequence a charged black hole quickly radiates away its charge. 

We conclude that in addition to the direct connection to the probability of emission in the high energy (charged) Hawking regime, the universal result for the tunneling integral also appears in the (modified) expression for the emission probability in the superradiant regime, which can be interpreted in terms of (generalized) Schwinger pair creation in the electric field near the horizon of the charged black hole. Indeed, for large black holes it was shown in \cite{Gibbons1975} that the emission of charged quanta is dominated by Schwinger pair production, rather than the Hawking process, and allows charged black holes to quickly get rid of their charge. The derived probability distribution generalizes that result by taking into account the backreaction which, as before, can be expressed in terms of the change of the black hole entropy. 

The superradiant regime and the inclusion of backreaction will play an important role in the next section. To be precise, so far we only considered non-extremal black holes for which the tunneling rate describes both thermal (neutral) radiation as well as charged (superradiance). Next, we will take the extremal limit of charged black holes. Since extremal black holes have a vanishing temperature, one expects the neutral (thermal) emission to shut down but charged decay channels should remain present. Clearly, to avoid creating a naked singularity only tunneling of (super-)extremal particles with $m \leq \omega \leq q$ is allowed. We will see that the tunneling calculation in the extremal limit not only confirms this expectation but in addition suggests the existence of a family of (non-extremal) gravitational instantons in the near-horizon $AdS_2 \times S^2$ limit, in which the superradiant regime is decoupled. 

\section{Extremal and near-horizon limits} \label{sec:extremal}

We would now like to study the universal result for charged emission from a charged black hole in the extremal limit. As is well known, the temperature of an extremal black hole vanishes, which is reflected by the fact that the emission rate \eqref{eq:decayrate} for neutral particles becomes zero in the extremal limit $M=Q$. However, we are interested to see what happens to the charged decay channels in the extremal limit. After a careful examination and regularization of the extremal limit, we will conclude that those decay channels are still captured by the universal expression for the tunneling integral. Once that has been established we will consider the near-horizon limit and relate the tunneling decay rate to gravitational instantons describing the spontaneous nucleation of domain walls in $AdS$.    

\subsection{Charged particle decay in the extremal limit}

A description in terms of particles tunneling out of an extremal black hole, for which the inner and outer horizon overlap, naively seems to be problematic due to the absence of a tunneling barrier lying in between the inner and outer horizon. The latter seems to be required to allow for a proper interpretation and related derivation of the final tunneling integral. One should be careful just extrapolating the final result, as it might be inconsistent and the different steps in the derivation need to be understood properly as one takes the extremal limit. 

To regularize the extremal limit, we will introduce a non-extremality parameter $\epsilon \ll Q$ defined as
\begin{align} \label{eq:extrsub}
r_+ &= Q + \epsilon ~, \nn\\
r_- &= Q - \epsilon ~,
\end{align}
such that the metric becomes
\begin{align}
ds^2 &= -f(r) dt^2 + f(r)^{-1}dr^2 + r^2d\Omega_2^2 ~,  \nn \\
f(r) &=\frac{(Q-r)^2}{r^2} - \frac{\epsilon^2}{r^2} ~.
\end{align}
The extremal limit is then defined as $\epsilon\to 0$. To study the region $r_-\leq r\leq r_+$ we follow \cite{Carroll2009} by introducing the coordinates
\be
r = Q - \epsilon \cos(\chi) \,, \qquad t = \frac{Q^2}{\epsilon} \psi ~.
\ee
In this region $\chi$ is a spacelike and $\psi$ a timelike coordinate. Using these coordinates the metric becomes
\begin{align}
ds^2 &= Q^2\left( -h(\chi)^2 d\chi^2  + \frac{\sin^2(\chi)}{h(\chi)^2}d\psi^2 + h(\chi)^2d\Omega_2^2 \right)~,  \nn \\
h(\chi) &= 1-\frac{\epsilon}{Q}\cos(\chi) ~.
\end{align}
The proper distance between $r_+$ and $r_-$ is given by
\be
\tau = Q \int_0^\pi d\chi~ h(\chi) = \pi Q ~,
\ee
which is independent of $\epsilon$. We conclude just as \cite{Carroll2009}, perhaps somewhat surprisingly, that even in the extremal limit $\epsilon \to 0$ there remains a finite proper distance between the inner and outer horizon.

Now we can continue as before and compute the tunneling integral for (super-)extremal shells from an extremal black hole. For an extremal shell we find (in Painlev\'e-Gullstrand coordinates)
\begin{align}
ds^2 &= -f(r)d t_p^2 + 2\sqrt{1-f(r)}dt_p dr + dr^2 + r^2 d\Omega_2 ~, \nn\\
f(r) &=\frac{(r-Q+q)^2}{r^2} - \frac{\epsilon^2}{r^2} ~.
\end{align}
The final result can be calculated by taking into account the substitutions \eqref{eq:extrsub} and in the end sending $\epsilon \to 0$. The result for the tunneling integral is
\be \label{eq:extremaltunnel}
\exp\left(-2~\im{ \int_{r_i}^{r_f} dr ~ p_c}\right) = e^{\pi\left( (Q-q)^2 - Q^2 \right)} = e^{\Delta S_{BH}} ~,
\ee
in full agreement with the universal expression. Similarly, we could also consider emission of super-extremal shells from an extremal black hole by making the substitutions
\begin{align}
M &\to Q - \omega ~,  \nn \\
Q &\to Q-q ~,
\end{align}
in the metric to describe a non-extremal black hole as the final state. On the other hand, if we were to consider sub-extremal shells, $r_+$ becomes imaginary after emission of the shell, which implies that the tunneling integral vanishes. Therefore, the emission of a sub-extremal particle (that would create a naked singularity) is forbidden.

We conclude that the same universal expression in terms of the black hole entropy difference still applies in the extremal limit. Although for an extremal black hole neutral emission shuts down, it can still decay via charged particles and the probability for that to happen can be expressed in terms of the (negative) entropy difference,  as anticipated. If we consider the emission of super-extremal shells satisfying 
\be
\omega + \frac{q^2}{2Q} < q ~,
\ee
we notice that the entropy difference becomes positive and therefore this process is governed by the superradiant expression for the probability that was derived previously. In contrast, we note that by including backreaction a parameter window opens up for shells satisfying 
\be
 q\left(1-\frac{q}{2Q}\right)< \omega \leq q ~,
\ee
that can be described by a (suppressed) tunneling amplitude, instead of the (lower energy) regime of superradiant emission. From a near-horizon point of view one might anticipate that these decay channels can be understood in terms of an instanton. In fact, in the near-horizon limit of a four-dimensional extremal Reissner-Nordstr\"om black hole \cite{Maldacena1999} derived the action for an instanton with charge equal to its tension connecting an initial $AdS_2 \times S^2$ spacetime with charge $Q$ to two $AdS_2 \times S^2$ with charge $Q_1$ and $Q_2$ while keeping the total charge $Q=Q_1+Q_2$ fixed. They related this to the instanton found by Brill \cite{Brill1992} resulting in the following decay rate
\be \label{eq:Brilldecay}
P \sim e^{-2\pi Q_1 Q_2} ~,
\ee
which coincides with \eqref{eq:extremaltunnel} in the limit $q \ll Q$, i.e. to leading order in the backreaction. In this extremal case this is appropriately described as fragmentation, since the two different vacua coexist peacefully and the domain wall separating them is flat and static. 

Similarly, super-extremal domain walls should be related to the emission of super-extremal shells for which $q (1-q/2Q) < \omega < q$. Such a shell necessarily expands due to its electromagnetic repulsion describing an instability of the extremal near-horizon $AdS$ geometry. Starting from an extremal black hole, for all $q$ and $\omega$ satisfying $M=Q \geq q\geq \omega > q(1-q/2Q)$ the entropy difference is negative describing an exponentially suppressed tunneling rate. In the near-horizon limit this should be related to a decay of $AdS$ space through the creation and subsequent expansion of a (super-)extremal domain wall. In the next section we will make this connection to domain walls in the near-horizon $AdS$ geometry explicit by using the near-horizon relation between the $AdS$ energy parameter $U$, which we will define in a moment, and the asymptotic Minkowski space energy parameter $\omega$.

\subsection{The near-horizon limit, domain walls and gravitational instantons}\label{subsec:DWandGI}

In order to relate the extremal black hole tunneling rate to a near-horizon $AdS$ instanton one needs to introduce the relevant near-horizon energy parameter, instead of the asymptotic Minkowski energy parameter $\omega$ that we have used so far. To derive an expression for the local energy density of the shell, let us reconsider the situation where an extremal black hole with charge $Q$ emits an extremal shell with charge $q$. Before emission, the metric is given by
\begin{align}
ds^2 &= f(r) dt^2 + f(r)^{-2}dr^2 + r^2 d\Omega_2^2 ~, \nn \\
f(r) &= \frac{(r-Q)^2}{r^2} ~,
\end{align}
and after emission the charge of the solution is reduced to $Q-q$. In order for these two geometries to be consistently joined together by the shell we need to satisfy Israel's junction conditions \cite{Israel1966}. We place the shell at some fixed position $r$ and label coordinates on the shell by $x^i$. In the thin-wall approximation the condition we have to satisfy is (working in units where $G_N=1/M_p^2=1$)
\be
8\pi S^i_j = (\Delta K)\delta^i_j  - \Delta K^i_j  ~.
\ee
Here $S^i_j$ is the surface energy-momentum tensor of the shell and $\Delta K_{ij}$ is the difference between extrinsic curvature on both sides of the shell. The energy density $\rho$ of a shell is then given by
\be \label{eq:localenergy}
\rho =  \frac1{8\pi}\left(\Delta K  - \Delta K^t_t \right) = \frac{1}{4\pi r}\left(\sqrt{f_-(r)}-\sqrt{f_+(r)}\right) ~,
\ee 
where $f_-(r)$ denotes the geometry with mass $M-\omega$ and charge $Q-q$ and $f_+(r)$ the geometry with $M$ and $Q$. The extremal $(\omega=q)$ shell has an energy density equal to
\be \label{eq:extshell}
\rho_{ext} = \frac{q}{4\pi r^2} ~.
\ee
If the shell can be viewed as a domain wall in the near-horizon limit, this energy density should be equal to the tension of an extremal domain wall. An extremal domain wall separating two (supersymmetric) vacua with vacuum energy $V_1$ and $V_2$ has a tension $T_{ext}$ that is given by \cite{Cvetic1997}
\be
8\pi T_{ext} = \frac{2}{\sqrt{3}}\left(\sqrt{|V_1|} - \sqrt{|V_2|}\right) \,,
\ee
where we take $|V_1| > |V_2|$. For two $AdS$ spaces of charge $Q_1=Q-q$ and $Q_2=Q$ the vacuum energy is given by $|V_1| = 3/(Q-q)^2$ and $|V_2| = 3/Q^2$. Thus, the tension of an extremal domain wall separating these two vacua is
\be \label{eq:exttension}
T_{ext} = \frac{q}{4\pi Q(Q-q)} = \frac{q}{4\pi Q^2} + {\cal O}\left(q^2/Q^2\right) \,,
\ee
where we assumed the probe limit $q\ll Q$. This indeed matches with the tension of an extremal shell, as given by \eqref{eq:extshell} in the near horizon limit $r \to Q$, provided $q\ll Q$. This confirms that extremal particle shells can be interpreted as flat, extremal domain walls from the point of view of the near-horizon geometry. 

Similarly, in the near-horizon limit super-extremal shells should correspond to super-extremal domain walls whose tension is bounded by $T<T_{ext}$. To make this correspondence explicit let us derive an expression for the near-horizon $AdS$ energy 
\be
U = r^2 \int d\Omega_2~ \rho ~,
\ee
which is a function of the asymptotic energy $\omega$ and the charges $Q$ and $q$. Here, $d\Omega_2$ is the volume element of the unit 2-sphere. Integrating the local energy density $\rho$ in the spherical shell, as given by \eqref{eq:localenergy}, and taking the near-horizon limit one derives
\be \label{eq:AdSEnergy}
U^2 = q^2- 2Q (q-\omega) ~.
\ee
Note that for shells satisfying $\omega < q$ this similarly implies $U<q$. This confirms that $T(\omega<q)<T_{ext}$ by recognizing that the tension can be expressed as $T=U/4\pi Q^2$.

Several additional comments are in order regarding the domain wall energy \eqref{eq:AdSEnergy}. In the extremal limit $\omega=q$ one indeed finds, as should be expected, that this implies $U=q$ as well. Inverting this relation gives $\omega = q(1-q/2Q) + U^2/2Q$ and as a consequence the near-horizon domain wall energy $U$ vanishes when the asymptotic energy is equal to $\omega=q(1-q/2Q)$, which exactly corresponds to the transition point where the entropy difference vanishes and the decay turns superradiant. We conclude that the near-horizon limit decouples this regime, in the sense that for all $U \geq 0$ the asymptotic energy $\omega$ is always in the regime where the decay channel is described by a suppressed tunneling amplitude in terms of the inherited entropy difference. We also note that the probe limit ($q/Q \ll 1$) necessarily implies $\omega \sim q$ and therefore is related to a near-extremal particle decay channel of the parent extremal black hole.  

To summarize the above, we explicitly related a family of domain wall instabilities of the near-horizon geometry to decay channels of the parent black hole. This seems to be a realization of an old conjecture made by Brill. He suggested that there should be an instanton that describes a single extremal Reissner-Nordstr\"om black hole splitting into two or more extremal black holes that agrees with the Brill instanton in the interior throat region \cite{Brill1992}. Work towards this goal was presented in \cite{Chung2012}, where an instanton was found describing the splitting of the throat region into two or more connected throat regions. According to that work the probability for that specific process is only \emph{half} the entropy difference. Our results suggest that the tunneling integral corresponds to the Lorentzian continuation of Brill's conjectured instanton. In fact, by taking backreaction into account the gravitational instanton related to the Hawking modes of a non-extremal black hole was first discussed in \cite{Massar1997}, where they indeed found a decay rate equal to  
\be \label{eq:tunnelinstanton}
P \sim e^{\Delta A_{BH}/4}  = e^{\Delta S_{BH}} \,,
\ee
in full agreement with the tunneling result (for $\Delta S_{BH} <0$). To extend their results to the extremal near-horizon limit we can regulate it as before using \eqref{eq:extrsub}, at the end sending $\epsilon\to 0$ and introducing $U$ as the relevant near-horizon energy parameter. For the special case where the extremal black hole emits an extremal shell, the instanton involved in this process should correspond to Brill's conjectured instanton. 

To be precise, we will now show explicitly that the expression for an extremal black hole emitting an extremal shell indeed reduces to the Brill instanton in the near-horizon limit. To do so, we continue the metric used in the tunneling calculation to Euclidean signature. Before we do this, it should be noted that Brill considered a magnetically charged solution whereas we are interested in an electrically charged solution. In order to obtain a real-valued instanton action, the electrical charge also has to be appropriately continued as $Q\to i Q$ \cite{Ng2002}. The Euclidean metric is then given by
\begin{align}
ds^2 &= f(r) dt^2 + \frac1{f(r)}dr^2 + r^2 d\Omega_2^2 \nn \\
f(r) &= \frac{(r-{\cal Q})^2}{r^2} \, .
\end{align}
Here we defined
\begin{align}
{\cal Q} = 
\begin{cases}
Q & (r<{\hat r}) ~,  \\
 \\
Q-q & (r>{\hat r})  \,,
 \end{cases}
\end{align}
where ${\hat r}$ denotes the position of the extremal shell. Both the geometry inside and outside the shell are Reissner-Nordstr\"om geometries, with a charge of respectively $Q-q$ and $Q$. We can therefore take the near horizon limit either inside or outside the shell. This limit is defined by writing
\be
r = {\cal Q} + \chi~ ,
\ee
and expanding around $\chi = 0$. The metric then becomes
\be
ds^2 = \frac{\chi^2}{{\cal Q}^2} dt^2 + \frac{{\cal Q}^2}{\chi^2} d\chi^2 + {\cal Q}^2 d\Omega_2^2 \,.
\ee
This can be rewritten in a form used by Brill
\begin{align} \label{eq:EuclidNH}
ds^2 &=H^2 dt^2 + H^{-2}\left(dx^2+dy^2+dz^2\right) ~,\nn \\
H&= \frac{{\cal Q}}{|\vec x|} ~,
\end{align}
where $|\vec x|^2 = x^2+y^2+z^2$. The Lorentzian version of this geometry, known as the Bertotti-Robinson geometry, corresponds to $AdS_2\times S^2$ and is an exact solution to the Einstein-Maxwell equations. More general, there also exist solutions with $N$ charges $Q_i$ for which
\be \label{eq:conformastatic}
H = \sum_{i=1}^N \frac{Q_i}{|\vec{x} - \vec x_i |} ~,
\ee
that are interpreted as a set of static extremal black holes with charge $Q_i$ placed at $\vec x_i$. 

We will now write down a particular two-centered black hole solution that agrees asymptotically with \eqref{eq:EuclidNH} by writing
\be
V=\frac{q}{|\vec x - \vec x_1|} + \frac{Q-q}{|\vec x|} \,.
\ee
As we will see, we can interpret this geometry as an extremal black hole placed at $\vec{x} = 0$ and our extremal shell placed at $\vec x_1$. It has the following asymptotic behavior 
\be \label{eq:BrillV}
\lim_{x \to \infty} V = \frac{Q}{|\vec x|} = H(r>\hat r) ~, \qquad \lim_{x \to 0} V =  \frac{Q-q}{|\vec x|} = H(r < \hat r)  ~.
\ee
We see that this geometry, at least asymptotically, agrees with \eqref{eq:EuclidNH}. Furthermore, we notice that \eqref{eq:BrillV} is precisely the Brill instanton where an $AdS_2\times S^2$ space of charge $Q$ splits into two $AdS_2\times S^2$ spaces with charge $Q-q$ and $q$. This leads us to the conclusion that the extremal near-horizon limit of the tunneling instanton found by \cite{Massar1998} for extremal shells is indeed the Brill instanton \cite{Brill1992}, as conjectured.

Whereas the Brill instanton describes the fragmentation of $AdS$ spaces, corresponding to the emission of an extremal shell from an extremal black hole, the tunneling expression should apply far more generally. In particular, it predicts that there should exist an entire family of gravitational instantons labeled by $\omega$ and $q$ that satisfy $q (1-q/2Q) < \omega \leq q$, for which the associated black hole entropy difference is always negative and therefore corresponds to a suppressed tunneling amplitude. From the perspective of the near-horizon limit these instantons describe the decay of $AdS$ vacua through the creation of super-extremal (expanding) domain walls connecting different vacua. The decay probability of these instantons should be provided by the tunneling integral for a finite window of parameters up until the extremal case. Indeed, the low-energy superradiant regime is decoupled in the near-horizon limit, as it would correspond to an imaginary near-horizon domain wall energy $U$. 

In fact, in the context of string theory some $AdS$ instantons of this type were already constructed in \cite{Maldacena1999}. There it was also observed that when the charge of a particle (0-brane) equals its tension, the Euclidean action of the corresponding instanton reduces to the value given by the Brill instanton. However, \cite{Maldacena1999} only derived this relation in the limit where the associated energy density (charge) of one of the $AdS$ spaces was small. Our results do not have such a restriction, as the tunneling integral is valid as long as $\omega \leq q \leq Q \leq M$. Nevertheless, in the limit where backreaction is small, our result should reduce to those of \cite{Maldacena1999}. By writing the tunneling amplitude for an extremal black hole emitting a shell with $\omega \leq q$ in terms of the domain wall energy $U$ we find
\be
\Delta S_{BH} = -2\pi Q\left(q - \sqrt{q^2-U^2} \right) + {\cal O}(U^2/Q^2) +  {\cal O}(q^2/Q^2) ~,
\ee
which as we already concluded reduces to the Brill instanton for $U=q$ and matches exactly with the $AdS_2$ instanton found for $U<q$ in \cite{Maldacena1999}. From the $AdS_2$ point of view, this super-extremal emission corresponds to Schwinger pair production \cite{Pioline:2005pf}. Since the superradiant regime is decoupled in the near-horizon limit, the general decay rate including backreaction in the spherically symmetric sector, should just be given by $P\sim e^{\Delta S_{BH}}$, extending the result of \cite{Maldacena1999} beyond the probe limit. 

To close this section, we conclude that the instabilities of (non-supersymmetric) $AdS$ space, as conjectured in an extension of the WGC in \cite{Ooguri2016,Freivogel2016}, are related to the (charged) decay channels that satisfy $q(1-q/2Q)<\omega < q$ of the parent extremal black hole geometry. The resulting decay rate can be expressed, including backreaction, in terms of the associated entropy difference. This result is not restricted to extremal shells, for which it was already noticed in \cite{Maldacena1999,Brill1992}, but extends to super-extremal shells.

\section{Conclusions and discussion} \label{sec:conclusion}

One of the original motivations to study backreaction corrections to the Hawking process was to potentially shed some light on how it could be consistent with unitary evolution of an underlying microscopic description. Treating the emitted particles as spherically symmetric shells and imposing energy conservation, Kraus and Wilczek derived an effective action for the shells and indeed found that the emission probability deviates from being exactly thermal \cite{Kraus1995a,Kraus1995}. In a similar spirit, Parikh and Wilczek imposed energy conservation to include backreaction by understanding the Hawking process in terms of a (spherically symmetric) quantum mechanical tunneling process. Their universal result \cite{Parikh2000} in terms of the difference of the black hole entropy before and after emission nicely supports a statistical thermodynamical interpretation of the transition, as was pointed out in earlier work by \cite{Massar1997}. One important conclusion of the current work is that the effective action approach reduces exactly to the tunneling integral. As a consequence both approaches are equivalent and the final result can always be expressed in terms of the entropy difference. This strongly suggests that the result can also be applied to describe the (spherically symmetric) decay of higher-dimensional black holes and/or black $p$-branes.   

As our prime example of interest we then derived the probability for emission of a charged shell from a charged black hole in terms of the exponential of the entropy difference of the black hole before and after emission. We then clarified the interpretation of the result in a low-energy regime of charged emission where the entropy difference changes sign and becomes positive. In this superradiant regime the probability for emission, including backreaction, has to be reassessed and we derived an expression that again features the entropy difference and reduces to the known expression for superradiant emission in the absence of backreaction. We then studied the extremal limit, showing that the backreacted result for charged (super-extremal) particle emission remains valid. The absence of decay channels for which $\omega\geq m>q$ is consistent with the weak cosmic censorship conjecture, and assuming the existence of super-extremal particles in the spectrum, as conjectured by the Weak Gravity Conjecture (WGC), the extremal black hole will decay. Another noteworthy result is that in the extremal black hole limit the inclusion of backreaction (and charge conservation) implies that the threshold energy below which superradiant behavior kicks in is distinguishably lower than the particle's charge $q$, opening a window of suppressed tunneling in the super-extremal emission regime. 

Having understood the extremal limit, corresponding to a near-horizon geometry of $AdS_2 \times S^2$, we then focused our attention on inherited decay channels of (non-supersym\-metric) $AdS$ spacetimes by identifying the relevant near-horizon $AdS$ energy. In the near-horizon limit a positive domain wall energy will always be in the suppressed (negative entropy difference) regime. Recently, the WGC conjecture was extended in \cite{Ooguri2016,Freivogel2016} by suggesting that the bound is only saturated for BPS states in a supersymmetric theory. This would imply that all non-supersymmetric $AdS$ spaces are unstable and will decay. In particular, \cite{Freivogel2016} motivated their conjecture by arguing that the WGC bounds the tension of (super-) extremal domain walls and therefore controls the stability of $AdS$ vacua. Extending the result for charged emission from charged black holes to the extremal near-horizon region, we indeed confirm that domain walls satisfying the WGC constraint will be spontaneously produced resulting in the decay of the $AdS$ geometry. The associated probability for this process is given by the universal expression in terms of the entropy difference of the parent black hole. Indeed, in the probe limit the tunneling amplitude exactly reproduces known results for super-extremal and extremal $AdS$ instantons.  

Strictly speaking our results only apply to $AdS_2$, but the universal form of the decay rate and its natural interpretation in terms of statistical thermodynamics suggests it applies equally well to higher-dimensional $AdS$ spacetimes, providing a very general and precise expression for the decay rate of $AdS$ through (super-) extremal domain walls, beyond the probe limit. It would of course be of interest to investigate this in more detail. In fact, higher-dimensional analogues of the Brill instanton that describe the fragmentation of higher-dimensional $AdS$ spaces claim to have been constructed in \cite{Ng2002}, seemingly at odds with general expectations from the AdS/CFT correspondence. We find our results also particularly intriguing in light of the work of \cite{Danielsson2017}. These authors claim that the instabilities they found for higher dimensional $AdS$ spaces are higher dimensional analogues of an instability discovered by Aretakis \cite{Aretakis:2011ha,Aretakis:2011hc}, which seems to be closely related to the onset of superradiance \cite{Zimmerman:2016qtn}. However, in our approach we noted that the superradiant particle decay channels of the parent black hole decouple in the near-horizon limit. Or, phrased differently, the would-be domain walls associated to superradiant decay of the parent black hole would have imaginary tension. Furthermore, in \cite{Danielsson2017} it is also argued that from the perspective of $AdS$ the instability is actually perturbative, signalled by open string modes becoming tachyonic. It would certainly be interesting to explore the relation between our semi-classical results and their top-down constructions further.

Let us finally emphasize once more that our results are expected to be universally valid in the spherically symmetric sector and can be applied whenever (super-) extremal particles are present in the low-energy effective action. As such, assuming the WGC holds, it should describe accurately the instabilities of charged black holes, as well as those of the corresponding near-horizon $AdS$ space in the extremal limit. Our findings therefore support the conjecture that all non-supersymmetric $AdS$ spaces are unstable and belong to the swampland, i.e. they cannot be consistently coupled to quantum gravity. What still remains to be more properly understood is how these generic results, derived from black hole physics, are related to specific (constraints on) potentials in low-energy effective field theory descriptions of $AdS$ spaces, and whether it allows for a generalization to include de Sitter spaces as well. We hope to come back to some of these questions in the near future.    

\section*{Acknowledgements}
We would like to thank Ben Freivogel, Bram van Overeem and Guilherme Leite Pimentel for useful discussions and Maulik Parikh and William Cottrell for comments on an early draft of this article. This work is part of the Delta ITP consortium, a program of the Netherlands Organisation for Scientific Research (NWO) that is funded by the Dutch Ministry of Education, Culture and Science (OCW). The work of LA and JPvdS is also supported by the research program of the Foundation for Fundamental Research on Matter (FOM), which is part of the Netherlands Organization for Scientific Research (NWO).

%%%
% Appendix
%%%

\appendix{

\section{Pole prescription and relevant integrals}
\label{app:integrals}

To calculate the rate of particles emitted by a charged black hole we need a pole prescription to evaluate the integrals \eqref{eq:KWint} and \eqref{eq:imaction}. In this appendix we show how the correct prescription is determined by the physical process under consideration and calculate the relevant integrals.

\subsection*{Pole prescription}
How to deal with the poles in the integrals we encountered can be understood by viewing the classical action as a propagator $K(x,x')$, which can be written as \cite{Hartle1976}
\be
K(x,x') = \sum_{paths} e^{iS(x,x')} ~,
\ee
where $S(x,x')$ is the classical action connecting the points $x$ and $x'$. Alternatively, the same propagator can also be viewed as a Green's function for the Klein-Gordon equation of a scalar field with mass $m$.
\be
(\nabla_\mu\nabla^\mu - m^2)K(x,x') = -\delta(x,x') ~.
\ee
Focussing on flat space for now, we write the propagator for a massless field in momentum space and obtain the well-known Feynman propagator
\be \label{eq:greenmom}
K_F(x,x') = \int\frac{d^4k}{(2\pi)^4}\frac{1}{k^2-i\epsilon} e^{ik(x-x')} ~,
\ee
which corresponds to deforming the contour as
\begin{align}
k^0 &= \omega_k - i\epsilon \qquad (\omega_k>0)~, \nn \\
k^0 &= \omega_k + i\epsilon \qquad (\omega_k<0) ~.
\end{align}
We can evaluate \eqref{eq:greenmom} by making use of a contour integral. If $t-t'>0$ we have to close the contour in the lower half-plane in order to be able to apply Jordan's lemma. This choice picks up the positive energy pole. We then obtain the well-known expression
\be
K_F(x,x') = -\frac{i}{4\pi^2} \frac{1}{s(x,x') + i \epsilon} ~,
\ee 
with $s(x,x')$ the square of the geodesic distance between $x$ and $x'$. Similarly, if $t-t'<0$ we have to close the contour in the upper half-plane which picks up the negative energy pole. We see that future directed propagation of positive energy particles corresponds to deforming the contour $\omega_k \to \omega_k - i \epsilon $ in momentum space and $t\to t - i\epsilon$ in position space.

Now we can use the results of \cite{Hartle1976} to obtain the analogous prescription for the Reissner-Nordstr\"om background. Also in this case, it was found that a future directed null geodesic corresponds to a deformation of the contour in the lower half $t$-plane, which in momentum space is equivalent to $\omega_k \to \omega_k - i \epsilon$, just as in flat space. We conclude that future propagation of positive energy particles requires $\omega_k \to \omega_k - i \epsilon $ and past propagation of negative energy particles $\omega_k \to \omega_k + i \epsilon$. This is also the prescription used in \cite{Parikh2000}.

\subsection*{Parikh-Wilczek integral}

The integral of interest is 
\be \label{eq:PWintegral}
S = -\int_{r_i}^{r_f} dr\int_0^{\omega} d\omega' ~\frac{1}{1-\sqrt{2(M-\omega')/r - (Q-q)^2/r^2 } - i\epsilon} ~,
\ee
where we used the prescription $\omega\to\omega-i\epsilon$. To calculate this integral, we first substitute $u=\sqrt{2(M-\omega)/r - (Q-q)^2/r^2}-1$ to find
\be \label{eq:uint}
S = -\int_{r_i}^{r_f} dr~r \int_{u(0)}^{u(\omega)} du~ \frac {u+1}{u + i\epsilon} = \int_{r_i}^{r_f} dr~r \int_{u(\omega)}^{u(0)} du~ \frac {u+1}{u + i\epsilon}  ~,
\ee
where in the right-hand side of the second equality we switched the boundaries of the integral because $u(0)>u(\omega)$. This integral has a pole at $u=-i\epsilon$. To evaluate it, we split it in terms of a principle value integral and a small (non-closed) contour that runs clockwise around the pole in the upper half-plane, as determined by the pole prescription. This contour is displayed in figure \ref{fig:contourPW}.
\begin{figure}[ht]
\centering
\includegraphics[scale=.8]{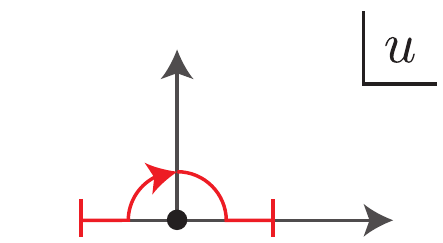}
\caption{Contour of the integral \eqref{eq:uint}, which has a pole at $u=-i\epsilon$.}
\label{fig:contourPW}
\end{figure}
The integral now becomes
\begin{align}
\lim_{\epsilon\to 0^+} \int_{u(\omega)}^{u(0)} du ~ \frac {u+1}{u + i\epsilon} &= {\cal P}\int_{u(\omega)}^{u(0)} du ~ \frac {u+1}{u} - i\pi \int_{u(\omega)}^{u(0)} du ~(u+1)\delta(u) \nn ~,\\
&={\cal P}\int_{u(\omega)}^{u(0)} du ~ \frac {u+1}{u} - i\pi ~.
\end{align}
Here, ${\cal P}$ denotes the principle value of the integral. It is straightforward to check that the principle value integral does not contribute an imaginary piece, and therefore we find
\be
\im{S}  = -\pi \int_{r_i}^{r_f} dr ~r = - \frac\pi 2(r_f^2-r_i^2) ~.
\ee

\subsection*{Kraus-Wilczek integral}
The second integral we need to evaluate appeared in section \ref{sec:UnivDecay}.
\be
I= \int_{r(0)}^{r(t)} dr ~ p_c  ~,
\ee
with $p_c$ given in \eqref{eq:canmom} and the boundaries by $r(0)=r_+(M-\omega,Q-q)-\epsilon$ and $r(t)=r_+(M,Q)+\epsilon$. Notice that to evaluate this integral, we could use Hamilton's equations to rewrite this integral in the form of \eqref{eq:PWintegral} to which we know the answer. Instead, for completeness and comparison with other references we will use the explicit expression of $p_c$.

Because  we are only interested in the imaginary part of this integral, we can focus on the logarithmic piece in $p_c$, since only this term can contribute an imaginary piece. Thus, the integral of interest is
\be
I = -\int_{r(0)}^{r(t)}dr~ r \log\left(\frac{ r - \sqrt{2r(M-\omega) - (Q-q)^2} -i\epsilon}{  r - \sqrt{2Mr - Q^2}}\right) ~.
\ee
This integral has branch points of the logarithm at $r_1 \equiv M-\omega+\sqrt{(M-\omega)^2-(Q-q)^2} + i \epsilon$ and $r_2 \equiv M+\sqrt{M^2-Q^2}$. Moreover, the argument of the logarithm has additional branch cuts at $r_3 \equiv (Q-q)^2/2(M-\omega)$ and  $r_4 \equiv Q^2/2M$. We choose the following branch cut structure.
\begin{align}
r_1 &: (-\infty,r_1] \qquad r_3: (-\infty,r_3] \nn \\
r_2 &: (-\infty,r_2] \qquad r_4: (-\infty,r_4] ~.
\end{align}
To evaluate $I$, we will split the integral in different parts.
\be \label{eq:KWcontour}
I = \lim_{\epsilon\to0^+}\left( \int_{C_L} + \int_{C^1_\epsilon} + \int_{C_1} + \int_{C^2_\epsilon} +  \int_{C_R} \right) dr~h(r) ~,
\ee
where
\be
h(r) = r \log\left(\frac{ r - \sqrt{2r(M-\omega) - (Q-q)^2} -i\epsilon}{  r - \sqrt{2Mr - Q^2}}\right) ~.
\ee
The structure of this integral in the complex plane is displayed in figure \ref{fig:contourKW}.
\begin{figure}[ht]
\centering
\includegraphics[scale=.8]{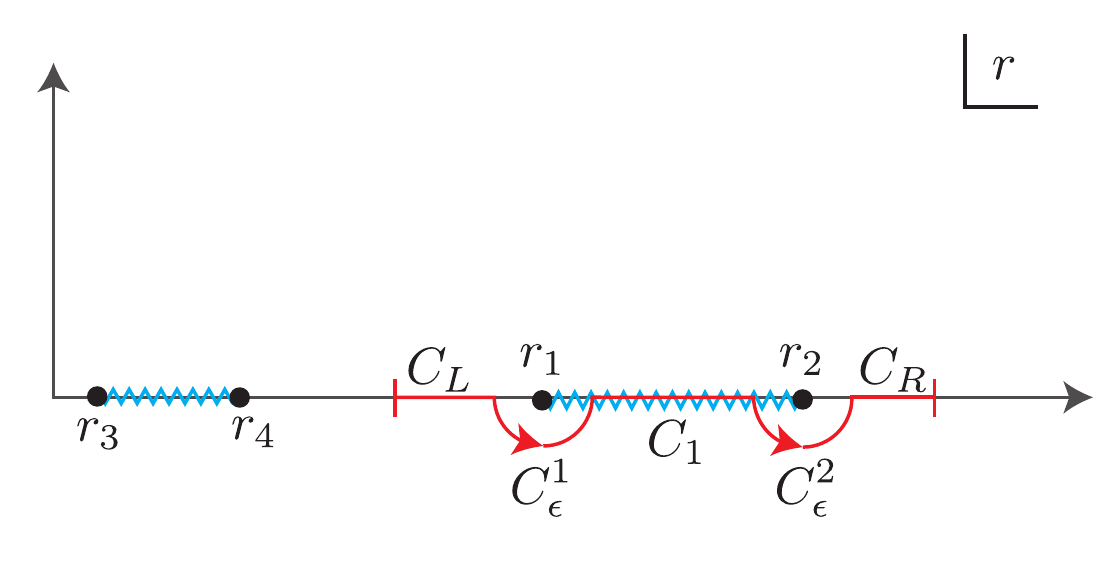}
\caption{Branch cut structure of the integral \eqref{eq:KWcontour}. This integral has two branch cuts that connect the points $r_1$ to $r_2$ and $r_3$ to $r_4$. The piece over $C_\epsilon^1$ and $C_\epsilon^2$ vanishes in the limit $\epsilon\to0$ and the parts over $C_L$ and $C_R$ are completely real.}
\label{fig:contourKW}
\end{figure}
When we approach the real axis between $r_1 <r < r_2$ from below, as required by the pole prescription, the logarithm takes the following form.
\be \label{eq:abovecut}
\log\left(\frac{ r - \sqrt{2r(M-\omega) - (Q-q)^2} }{r - \sqrt{2Mr - Q^2}}\right) = \log\left|\frac{ r - \sqrt{2r(M-\omega) - (Q-q)^2} }{  r - \sqrt{2Mr - Q^2}}\right| +i\pi
\ee
On the other hand, the imaginary part of the logarithm is zero on $C_L$ and $C_R$. Because we are only interested in the imaginary part of $I$, we can ignore these pieces. Finally, by observing that the integrals over $C_\epsilon^1$ and $C_\epsilon^2$ vanish in the limit $\epsilon \to 0$, we obtain
\be
\im{I} = - \pi \int_{r_+(M_-,Q_-)}^{r_+(M,Q)}dr~r = -\frac\pi2\left(r_+(M,Q)^2 - r_+(M_-,Q_-)^2\right) ~.
\ee

}

%%%
% Bibliography
%%%


\begin{thebibliography}{10}

\bibitem{Kraus1995a}
P.~Kraus and F.~Wilczek, \emph{{Self-interaction correction to black hole
  radiance}},
  \href{http://dx.doi.org/10.1016/0550-3213(94)00411-7}{\emph{Nucl. Phys. B} {\bfseries433} (1995) 403--420},
  [\href{https://arxiv.org/abs/gr-qc/9408003}{{\ttfamily gr-qc/9408003}}].

\bibitem{Kraus1995}
P.~Kraus and F.~Wilczek, \emph{{Effect of self-interaction on charged black
  hole radiance}},
  \href{http://dx.doi.org/10.1016/0550-3213(94)00588-6}{\emph{Nucl. Phys. B} {\bfseries 437} (1995) 231--242},
  [\href{https://arxiv.org/abs/hep-th/9411219}{{\ttfamily hep-th/9411219}}].

\bibitem{Parikh2000}
M.~K. Parikh and F.~Wilczek, \emph{{Hawking radiation as tunneling}},
  \href{http://dx.doi.org/10.1103/PhysRevLett.85.5042}{\emph{Phys. Rev. Lett.} {\bfseries 85} (2000) 5042--5045},
  [\href{https://arxiv.org/abs/hep-th/9907001}{{\ttfamily hep-th/9907001}}].
  
\bibitem{Massar1997}
S.~Massar and R.~Parentani, \emph{{Gravitational instanton for black hole
  radiation}},
  \href{http://dx.doi.org/10.1103/PhysRevLett.78.3810}{\emph{Phys. Rev. Lett.} {\bfseries 78} (1997) 3810--3813},
  [\href{https://arxiv.org/abs/gr-qc/9701015}{{\ttfamily gr-qc/9701015}}].

\bibitem{Bekenstein:1975tw}
J.~D. Bekenstein, \emph{{Statistical black-hole thermodynamics}},
  \href{http://dx.doi.org/10.1103/PhysRevD.12.3077}{\emph{Phys. Rev. D}
  {\bfseries 12} (1975) 3077--3085}.

\bibitem{Keski-Vakkuri1997}
E.~Keski-Vakkuri and P.~Kraus, \emph{{Microcanonical D-branes and back
  reaction}},
  \href{http://dx.doi.org/10.1016/S0550-3213(97)00085-0}{\emph{Nucl. Phys. B}
  {\bfseries 491} (1997) 249--262},
  [\href{https://arxiv.org/abs/hep-th/9610045}{{\ttfamily hep-th/9610045}}].
  
\bibitem{Arkani-Hamed2007}
N.~Arkani-Hamed, L.~Motl, A.~Nicolis and C.~Vafa, \emph{{The string landscape,
  black holes and gravity as the weakest force}},
  \href{http://dx.doi.org/10.1088/1126-6708/2007/06/060}{\emph{JHEP} {\bfseries
  06} (2007) 60},
  [\href{https://arxiv.org/abs/hep-th/0601001}{{\ttfamily hep-th/0601001}}].

\bibitem{Gibbons1975}
G.~W. Gibbons, \emph{{Vacuum polarization and the spontaneous loss of charge by
  black holes}},
  \href{http://dx.doi.org/10.1007/BF01609829}{\emph{Commun. Math. Phys.} {\bfseries 44} (1975) 245--264}.

\bibitem{Ooguri2016}
H.~Ooguri and C.~Vafa, \emph{{Non-supersymmetric AdS and the Swampland}},
  \href{https://arxiv.org/abs/1610.01533}{{\ttfamily 1610.01533}}.

\bibitem{Freivogel2016}
B.~Freivogel and M.~Kleban, \emph{{Vacua Morghulis}},
  \href{https://arxiv.org/abs/1610.04564}{{\ttfamily 1610.04564}}.

\bibitem{Maldacena1999}
J.~Maldacena, J.~Michelson and A.~Strominger, \emph{{Anti-de Sitter
  Fragmentation}},
  \href{http://dx.doi.org/10.1088/1126-6708/1999/02/011}{\emph{JHEP} {\bfseries
  02} (1999) 11}, [\href{https://arxiv.org/abs/hep-th/9812073}{{\ttfamily
  hep-th/9812073}}].

\bibitem{Brill1992}
D.~R. Brill, \emph{{Splitting of an extremal Reissner-Nordstr{\"{o}}m throat
  via quantum tunneling}},
  \href{http://dx.doi.org/10.1103/PhysRevD.46.1560}{\emph{Phys. Rev. D}
  {\bfseries 46} (1992) 1560--1565}.

\bibitem{painleve1921cr}
P.~Painlev{\'{e}}, \emph{{La m{\'{e}}canique classique et la th{\'{e}}orie de
  la relativit{\'{e}}}}, \href{http://adsabs.harvard.edu/abs/1922LAstr..36....6P}{{\emph{C. R. Acad. Sci. (Paris)} {\bfseries 173}
  (1921) 677--680}}.

\bibitem{Hawking1996}
S.~W. Hawking and G.~T. Horowitz, \emph{{The gravitational Hamiltonian, action,
  entropy and surface terms}},
  \href{http://dx.doi.org/10.1088/0264-9381/13/6/017}{\emph{Class. Quant. Grav.} {\bfseries 13} (1996) 1487--1498},
  [\href{https://arxiv.org/abs/gr-qc/9501014}{{\ttfamily gr-qc/9501014}}].

\bibitem{Massar1998}
S.~Massar and R.~Parentani, \emph{{On the Gravitational Back Reaction to
  Hawking Radiation}},  \href{https://arxiv.org/abs/gr-qc/9801043}{{\ttfamily
  gr-qc/9801043}}.

\bibitem{Hansen1981}
A.~Hansen and F.~Ravndal, \emph{{Klein's paradox and its resolution}},
  \href{http://dx.doi.org/10.1088/0031-8949/23/6/002}{\emph{Phys. Scripta.}
  {\bfseries 23} (1981) 1036--1042}.

\bibitem{Brito2015}
R.~Brito, V.~Cardoso and P.~Pani, \emph{{Superradiance}},
\emph{Springer} (2015).

\bibitem{Bekenstein:1977mv}
J.~D. Bekenstein and A.~Meisels, \emph{{Einstein $A$ and $B$ coefficients for a
  black hole}}, \href{http://dx.doi.org/10.1103/PhysRevD.15.2775}{\emph{Phys.
  Rev. D} {\bfseries 15} (1977) 2775--2781}.

\bibitem{Carroll2009}
S.~M. Carroll, M.~C. Johnson and L.~Randall, \emph{{Extremal limits and black
  hole entropy}},
  \href{http://dx.doi.org/10.1088/1126-6708/2009/11/109}{\emph{JHEP} {\bfseries
  11} (2009) 109}, [\href{https://arxiv.org/abs/0901.0931}{{\ttfamily
  0901.0931}}].
  
\bibitem{Israel1966}
W.~Israel, \emph{{Singular hypersurfaces and thin shells in general
  relativity}}, \href{http://dx.doi.org/10.1007/BF02710419}{\emph{Nuovo Cimento B} {\bfseries 44} (1966) 1--14}.

\bibitem{Cvetic1997}
M.~Cveti{\v{c}} and H.~H. Soleng, \emph{{Supergravity domain walls}},
  \href{http://dx.doi.org/10.1016/S0370-1573(96)00035-X}{\emph{Phys. Rep.}
  {\bfseries 282} (1997) 159--223},
  [\href{https://arxiv.org/abs/hep-th/9604090}{{\ttfamily hep-th/9604090}}].
  
\bibitem{Chung2012}
H.~Chung, \emph{Tunneling between single- and multicentered black hole
  configurations},
  \href{http://dx.doi.org/10.1103/PhysRevD.86.064036}{\emph{Phys. Rev. D}
  {\bfseries 86} (2012) 064036},
  [\href{https://arxiv.org/abs/1201.3028}{{\ttfamily
  1201.3028}}].
  
\bibitem{Ng2002}
S.~Ng and M.~Perry, \emph{{Brane splitting via quantum tunneling}},
  \href{http://dx.doi.org/10.1016/S0550-3213(02)00346-2}{\emph{Nucl. Phys. B}
  {\bfseries 634} (2002) 209--229},
  [\href{https://arxiv.org/abs/hep-th/0106008}{{\ttfamily hep-th/0106008}}].
  
\bibitem{Pioline:2005pf}
B.~Pioline and J.~Troost, \emph{{Schwinger pair production in $AdS_2$}},
  \href{http://dx.doi.org/10.1088/1126-6708/2005/03/043}{\emph{JHEP} {\bfseries
  03} (2005) 043}, [\href{https://arxiv.org/abs/hep-th/0501169}{{\ttfamily
  hep-th/0501169}}].

\bibitem{Danielsson2017}
U.~H. Danielsson, G.~Dibitetto and S.~C. Vargas, \emph{{A swamp of non-SUSY
  vacua}},
  \href{https://doi.org/10.1007/JHEP11(2017)152}{\emph{JHEP} {\bfseries
  11} (2017) 152}, [\href{https://arxiv.org/abs/1708.03293}{{\ttfamily
  1708.03293}}].
  
\bibitem{Aretakis:2011ha}
S.~Aretakis, \emph{{Stability and Instability of Extreme Reissner-Nordstr\"om
  Black Hole Spacetimes for Linear Scalar Perturbations I}},
  \href{http://dx.doi.org/10.1007/s00220-011-1254-5}{\emph{Commun. Math. Phys.}
  {\bfseries 307} (2011) 17--63},
  [\href{https://arxiv.org/abs/1110.2007}{{\ttfamily 1110.2007}}].

\bibitem{Aretakis:2011hc}
S.~Aretakis, \emph{{Stability and Instability of Extreme Reissner-Nordstr\"om
  Black Hole Spacetimes for Linear Scalar Perturbations II}},
  \href{http://dx.doi.org/10.1007/s00023-011-0110-7}{\emph{Ann. Henri Poincar\'e}
   {\bfseries 12} (2011) 1491--1538},
  [\href{https://arxiv.org/abs/1110.2009}{{\ttfamily 1110.2009}}].

\bibitem{Zimmerman:2016qtn}
P.~Zimmerman, \emph{{Horizon instability of extremal Reissner-Nordstr\"om black
  holes to charged perturbations}},
  \href{http://dx.doi.org/10.1103/PhysRevD.95.124032}{\emph{Phys. Rev.}
  {\bfseries D95} (2017) 124032},
  [\href{https://arxiv.org/abs/1612.03172}{{\ttfamily 1612.03172}}].

\bibitem{Hartle1976}
J.~B. Hartle and S.~W. Hawking, \emph{{Path-integral derivation of black-hole
  radiance}}, \href{http://dx.doi.org/10.1103/PhysRevD.13.2188}{\emph{Phys.
  Rev. D} {\bfseries 13} (1976) 2188--2203}.

\end{thebibliography}
\end{document}